\documentclass[sigplan,screen,10pt]{acmart}

\usepackage{cleveref} 
\crefformat{section}{\S#2#1#3}
\crefformat{subsection}{\S#2#1#3}
\crefformat{subsubsection}{\S#2#1#3}

\usepackage{pifont}
\usepackage{subfigure}
\usepackage{multirow}
\usepackage{caption}

\AtBeginDocument{%
  \providecommand\BibTeX{{%
    \normalfont B\kern-0.5em{\scshape i\kern-0.25em b}\kern-0.8em\TeX}}}

\setcopyright{none}
\renewcommand\footnotetextcopyrightpermission[1]{}
\begin{document}

\title{Revisiting Swapping in User-space with Lightweight Threading}


\author{Kan Zhong$^\dagger$, Wenlin Cui$^\dagger$, Youyou Lu$^\S$, Quanzhang Liu$^\dagger$, \\ Xiaodan Yan$^\dagger$, Qizhao Yuan$^\dagger$, Siwei Luo$^\dagger$, and Keji Huang$^\dagger$}
\affiliation{%
	\institution{$^\dagger$Huawei Technologies Co., Ltd} 
	\city{Chengdu}
	\country{China}
}
\affiliation{%
	\institution{$^\S$Department of Computer Science and Technology, Tsinghua University}
	\city{Beijing}
	\country{China}
}

\renewcommand{\shortauthors}{XXX, et al.}

\begin{abstract}
Memory-intensive applications, such as in-memory databases, caching systems and key-value stores, are increasingly demanding larger main memory to fit their working sets. Conventional swapping can enlarge the memory capacity by paging out inactive pages to disks. However, the heavy I/O stack makes the traditional kernel-based swapping suffers from several critical performance issues.

In this paper, We redesign the swapping system and propose \textit{\textbf{LightSwap}}, an high-performance user-space swapping scheme that supports paging with both local SSDs and remote memories. First, to avoids kernel-involving, a novel page fault handling mechanism is proposed to handle page faults in user-space and further eliminates the heavy I/O stack with the help of user-space I/O drivers. Second, we co-design Lightswap with light weight thread (LWT) to improve system throughput and make it be transparent to user applications. Finally, we propose a try-catch framework in Lightswap to deal with paging errors which are exacerbated by the scaling in process technology.

We implement Lightswap in our production-level system and evaluate it with YCSB workloads running on memcached. Results show that Ligthswap reduces the page faults handling latency by 3--5 times, and improves the throughput of memcached by more than 40\% compared with the stat-of-art swapping systems.


\end{abstract}



\maketitle
\pagestyle{plain}

\section{Introduction}

Memory-intensive applications\cite{memcached, voltdb}, such as in-memory databases, caching systems, in-memory key-value stores are increasingly demanding more and more memory to meet their low-latency and high-throughput requirements as these applications often experience significant performance loss once their working set cannot fit in memory. Therefore, extending the memory capacity becomes a mission-critical task for both researchers and system administrators.

Based on the virtual memory system, existing OS provides swapping to enlarge the main memory by writing inactive pages to a backend store, which today is usually backed by SSDs. Compared to SSDs, DRAM still has orders of magnitude performance advantage, providing memory-like performance by paging with SSDs has been explored for decades~\cite{FlashVM, new-linux-swap, CFLRU, os-for-hybrid, FlatFlash, speculative-paging, flashmap, mmio-flash, SSDAlloc, NVMalloc, ssd-hybird-memcached} and still remains great challenges. Especially, we find that the heavy kernel I/O stack introduces large performance penalty, more than 40\% of the time is cost by the I/O stack when swapping in/out a signal page, and this number will keep increasing if ultra-low latency storage media, such as Intel Optane~\cite{intel-optane} and KIOXIA XL-FLash~\cite{XL-Flash} are adopted as the backend stores.

To avoid the high-latency of paging with SSDs, memory disaggregation architecture~\cite{res-disagg, net-support, disagg-mem, acc-db, mem-disagg-flx} proposes to expose a global memory address space to all machines to improve memory utilization and avoid memory over-provisioning. However, existing memory disaggregation proposals require new hardware supports~\cite{hp-the-machine, net-support, disagg-mem, sys-implication, rethink-disagg-mem}, making these solutions infeasible and expensive in real production environments. Fortunately, recent researches, such as Infiniswap~\cite{infiniswap}, Fluidmem~\cite{FluidMem}, and AIFM~\cite{AIFM} have shown that paging or swapping with remote memory is a promising solution to enable memory disaggregation. However, we still find several critical issues of these existing approaches.

First, kernel-based swapping, which relies on the heavy kernel I/O stack exhibits large software stack overheads, making it cannot fully exploit the high-performance and low-latency characteristics of emerging storage media (e.g., Intel Optane) or networks (e.g., RDMA). We measured the remote memory access latency of Infiniswap, which is based the kernel swapping, can be as high as 40$\mu$s even using one-side RDMA. Further, modern applications exhibit diverse memory access patterns, kernel-based swapping fails to provide enough flexibility to customize the eviction and prefetching policy. Second, recent research~\cite{FluidMem} has also explored the user-space swapping with userfaultfd, which however needs extra memory copy operations and exhibits ultra-high page fault latency under high-concurrency (i.e., 107$\mu$s under 64 threads), leading to systems that based on userfaultfd cannot tolerate frequent page faults. Finally, new proposals like AIFM~\cite{AIFM} and Semeru~\cite{Semeru}, that do not rely on virtual memory abstraction, provide a runtime-managed swapping to applications and can largely reduce the I/O amplification. However, these schemes break the compatibility and require large efforts to rewrite existing applications.

Therefore, we argue that swapping need to be redesigned to become high-performance and remains transparent to applications. In this paper, we propose \emph{Lightswap}, an user-space swapping that supports paging with both SSDs and remote memories. First, Lightswap handles page faults in user-space and utilizes the high-performance user I/O stack to eliminate the software stack overheads. To this end, we propose an ultra-low latency page fault handling mechanism to handle page faults in user-space (\cref{sec4-2}). Second, when page faults happen, existing swapping schemes will block the faulting thread to wait for data fetch, which lowers the system throughput. In Lightswap, we co-design swapping with light weight thread (LWT) to achieve both high-performance and application-transparent (\cref{sec4-3}). When page fault happens, leveraging the low context switch cost of LWT, we use a dedicated swap-in LWT to handle page fault and allow other LWTs to occupy the CPU while waiting for data fetch. Finally, due to the scaling in process technology and the ever increasing capacity, we find that both DRAM and SSD become more prone to errors. Therefore, we propose a try-catch exception framework in Lightswap to deal with paging errors (\cref{sec4-4}). 

We evaluate Lightswap with memcached workloads. Evaluation results show that Lightswap can reduce the page fault handling latency by 3--5 times and improve the throughput by more than 40\% when compared to Infiniswap. In summary, we make the following contributions:
\vspace{-0.05in}
\begin{itemize}
\item We propose an user-space page fault handling mechanism that can achieve ultra-low page fault notification latency (i.e., 2.4$\mu$s under 64 threads).
\item We demonstrate that with the help of LWT, user-space swapping system can achieve both high-performance and application-transparent.
\item We propose a try-catch based exception framework to handle paging errors. To best of our knowledge, we are the first work to explore handling both memory errors and storage device errors in a swapping system.
\item We show the performance benefits of Lightswap with YSCB workloads on memcached, and compare it to other swapping schemes.
\end{itemize}

The rest of the paper is organized as follows.  Section~\ref{sec:2} presents the background and motivation. Section~\ref{sec:3} and \ref{sec:4} discuss our design considerations and the design details of Lightswap, respectively. Section~\ref{sec:5} presents the implementation and evaluation of Lightswap. We cover the related work in Section~\ref{sec:6} and conclude the paper in Section~\ref{sec:7}.


\section{Background and Motivation}
\label{sec:2}

\subsection{Swapping}

\begin{figure}[t]
	\centering
	\includegraphics[width=0.95\columnwidth]{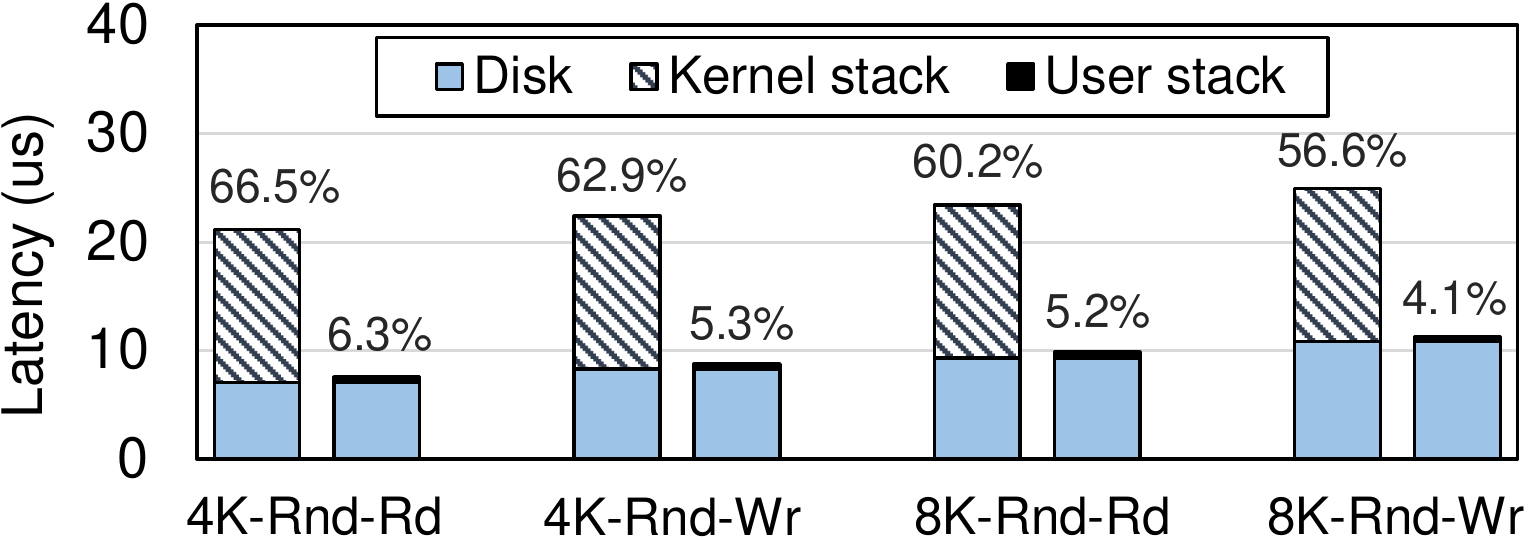} \\
	\caption{Random read and write latency breakdown. \textit{The numbers on top of each bar denote the relative fraction of I/O stack time in the total latency.}}
	\label{fig: io-stack-lat}
	\vspace{-0.15in}
\end{figure}



Existing swapping approaches~\cite{FlashVM, new-linux-swap, CFLRU, os-for-hybrid, FlatFlash, speculative-paging, flashmap, mmio-flash} depend on the existing kernel data path that is optimized for slow block devices, both reading and writing pages from/to the backend store would introduce high software stack overheads. Figure~\ref{fig: io-stack-lat} compares the I/O latency breakdown for random read and write of XL-Flash~\cite{XL-Flash} while using the kernel data path and user-space SPDK~\cite{spdk} driver. The figure shows that over half (56.6\% -- 66.5\%) of the time is spent on the kernel I/O stack for both read and write while using the kernel data path. This overhead mostly comes from the generic block layer and device driver. Comparably, the I/O stack overhead is negligible while using the user-space SPDK driver.

Besides using local SSDs as the backend store, the high bandwidth and low latency RDMA network offers the opportunity for swapping pages to remote memories. The newly proposed memory disaggregation architecture~\cite{res-disagg, net-support, disagg-mem, acc-db, mem-disagg-flx} takes a first look on remote memory paging. In disaggregation model, computing nodes can be composed of large amount of memory borrowing space from remote memory servers. Existing works, such as Infiniswap~\cite{infiniswap} and FluidMem~\cite{FluidMem} have showed that swapping to remote memories is a promising solution for memory disaggregation. Nevertheless, Infniswap exposes remote memory as a local block device, paging in/out still needs to go through the whole kernel I/O stack. Our evaluation shows that the remote memory access in Infiniswap can as high as 40$\mu$s, which is about 10x higher than the latency of a 4KB page RDMA read. The difference is all caused by the slow kernel data path.

\begin{figure}
	\centering
	\includegraphics[width=0.8\columnwidth]{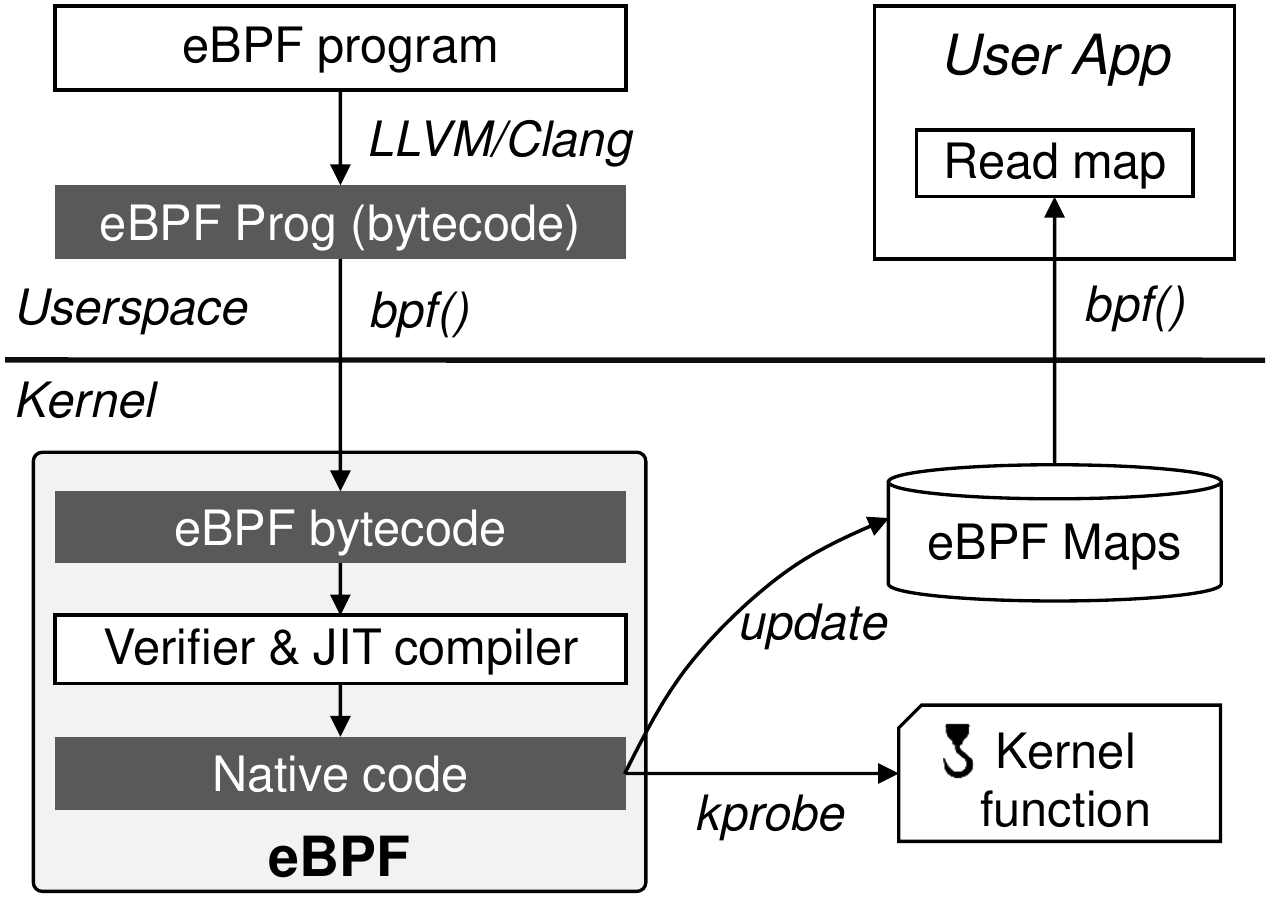} \\
	\caption{Linux eBPF framework.}
	\label{fig: ebpf}
	\vspace{-0.15in}
\end{figure}

\subsection{Linux eBPF}


eBPF (for \textit{extended Berkeley Packet Filter}) is a general virtual machine that running inside the Linux kernel. It provides an instruction set and an execution environment to run eBPF programs in kernel. Thus, user-space applications can instrument the kernel by eBPF programs without changing kernel source code or loading kernel modules. eBFP programs are written in a special assembly language. Figure~\ref{fig: ebpf} shows how eBPF works. As shown, eBPF bytecode can be loaded into the kernel using \texttt{bpf()} system call. Then a number of checks are performed on the eBPF bytecode by a verifier to ensure that it cannot threaten the stability and security of the kernel. Finally, the eBPF bytecode can either be executed in the kernel by an interpreter or translated to native machine code using a Just-in-Time (JIT) compiler.


eBPF programs can be attached to predetermined hooks in the kernel, such as the traffic classifier (\texttt{tc})~\cite{tc} and eXpress Data Path (XDP)~\cite{xdp} in the network stack. One can also attach eBPF programs to \texttt{kprobe} tracepoint hooks, which makes eBPF programs can be attached to any kernel function. One of most important property of eBPF is that it provides \textit{eBPF maps} for sharing data with user-space application. eBPF maps are data structures implemented in the kernel as key-value stores. Keys and values are treated as binary blobs, allowing to store user-defined data structures and types. To enable handling page fault in user-space, we utilize eBPF program to store thread context and page fault information into maps when page fault happens. Then our user-space page fault handler can retrieve the needed information from the maps to handle the page fault (\cref{sec4-2}).


\subsection{Target Application and Execution Mode}
In high-concurrency and low-latency memory-intensive applications, such as web service, in-memory cache and database, each server may handle thousands of requests. The traditional thread-based execution model (i.e., launching one thread per request) would lead to significant scheduling overhead, making most of the CPU time wasted on thread scheduling and context switching. To address this issue, light weight thread (LWT), as known as coroutine~\cite{coroutine, revis-crt} is proposed and widely adopted in memory-intensive applications to improve the throughput and reduce the response latency~\cite{expl-crt, interlv-crt}. Different from thread, such \emph{pthread} in Linux, LWT is fully controlled and scheduled by user program, instead of the operating system. Each thread can be comprised of lots of LWTs, and each LWT has an entry function that can suspend its execution and resume at a later point. Therefore, compared to OS managed thread, LWT has much lower scheduling overhead and more flexible scheduling policy.

Moreover, in conventional swapping, when accessing a non-present page, the current thread will be blocked and woke up by the OS once the requested page is fetched back to local DRAM. During this process, the swap-in request needs to go through the whole IO stack to read the page, and a context switch is performed to put the blocked thread into running state, which both bring significant latency penalty. 

Therefore, we adopt LWT as our application execution model for high-throughput in-memory systems, and co-design Lightswap with LWT to provide an high-performance and transparent user-space swapping to applications.

\section{\emph{LightSwap} Design Considerations}
\label{sec:3}

This section discusses the design considerations of Lightswap. To make Lightswap fast and flexible, we move page swapping from kernel to user-space, and co-design swapping with LWT to hide the context switching cost and improve the CPU utilization. Then, we discuss the need of paging error handling due to the scaling in process technology.

\subsection{Why User space?}
\label{sec3-1}
We design an user space swapping framework based on the following reasons:

\textit{ \textbf{1) User space I/O drivers show high potential in performance improvements.}} The performance of storage devices has been improved significantly due to the emerging technologies in both storage media and interface, such the Intel Optane memory~\cite{intel-optane}, new NVMe (Non-Volatile Memory Express)~\cite{nvme} interface and PCI Express (PCIe) 4.0 interconnect. Therefore, the overhead of legacy kernel I/O stack becomes more and more noticeable since it was originally optimized for slow HDDs. 

To reduce the I/O stack overhead, user space I/O stacks without any kernel intervention are desired to support high-performance storage device. To this end, Intel released storage performance development kit (SPDK)~\cite{spdk}, which moves all the necessary drivers into user-space, thus avoids syscalls and enables zero-copy. Other optimizations, such as polling and lock-free are also used in SPDK to enhance the I/O performance. To accelerate network I/Os, DPDK~\cite{dpdk}, a packet process acceleration framework, maps Ethernet NIC into user-space and control it in user-space program. DPDK also provides an user-space polling mode driver (PMD), which uses polling to check for received data packets instead of using interrupts as the kernel network stack would. Therefore, to be beneficial from these high-performance user-space I/O drivers, we build Lightswap in user-space.

\textit{ \textbf{2) User space swapping can easily support memory disaggregation.}} Thanks to the fast and low-latency RDMA network, the effective memory capacity can also be extended through remote memory paging. To achieve this, memory disaggregation architecture~\cite{res-disagg, net-support, disagg-mem, acc-db, mem-disagg-flx} has been proposed to expose the memories in dedicated servers to computational severs to enlarge their memory space. Previous works~\cite{infiniswap, FluidMem, xmempod, soft-far-memroy} have shown that swapping is a promising solution to enable efficient memory disaggregation.

Lightswap uses an user-space key-value abstraction for paging in/out. Pages are read/write from/to backend stores through an unified KV interface. Thus, memory disaggregation can be enabled by writing pages to a remote KV store in Lightswap. 



\textit{ \textbf{3) User-space is more resilient to errors.}} Due to the continuous scale in storage density and process technology, both storage devices and memories becomes more prone to errors~\cite{flash-error}. When these errors are triggered in kernel space, Linux and UNIX-like OSes have to call a panic function, which cause system crash and even service termination.

To deal with the above errors in user-space, one can isolate the faulted storage device or memory address and then simply kill the corresponding applications. However, this approach still causes application termination and thus lower the system's availability. Moreover, it is difficult to handle the error properly without application semantics. Therefore, we propose to handle memory and device errors in user-space with application-specific knowledges.





\subsection{Swapping with LWT}
Existing OS usually pays a high cost on thread context-switching, which can take 5--10 microseconds on x86 platforms~\cite{linux-perf}. To hide the thread context-switching latency, Fastswap~\cite{far-memory} poll waits the requested page when page fault happens by leveraging the low latency of RDMA network. However, with SSD-based swapping or larger page size, reading the requested page into local memory needs comparably longer durations. Even paging with remote memory with RDMA, we still argue that polling wait is not the optimal way as the context-switching of LWT is in nanoseconds. Thus, polling wait would cause a large waste on CPU cycles. To tackle this issue, Lightswap uses asynchronous I/O: it switches to other LWTs while waiting for data fetch.

To effectively swap-in pages in user-space, we co-design Lightswap with LWT (\cref{sec4-3}). First, to make Lightswap be transparent to applications, Lightswap uses a dedicated LWT, referred as \textit{swap-in LWT}, to fetch pages from a backend store to local memory. When a page fault is triggered by normal application LWT, referred as \textit{faulting LWT}, it will be blocked and the swap-in LWT is launched to fetch the requested pages. Second, the swap-in LWT will also be blocked and yields CPU for other worker LWTs when waiting for data fetch. Finally, to reduce the overall page fault latency, we adjust our LWT scheduler to prioritize swap-in LWTs, thus the requested pages can be fetched as soon as possible.


\subsection{Handling Paging Errors}

Due to storage device and memory errors, pages in SSD-based backend store, remote memory, and local memory have more possibility to be corrupted. Therefore, paging mainly encounter two kinds of errors: 1) \textit{swap-in error}, swapped out pages cannot be brought back due to device or network failure, and 2) \textit{uncorrectable memory error (UCE)}, memory error that has exceeded the correction capability of DRAM hardware. Existing data protection methods, such as replication and erasure code, only work well for slow disk. To reduce the possibility of encounter memory errors in memory access, a daemon named \texttt{memory scrubber} will periodically scan DRAM and correct any potential errors. However, the most advanced DRAM ECC scheme (i.e., chipkill) also fails to correct errors from multiple devices in a DIMM module~\cite{dram-err}. When an UCE is found by memory scrubber or triggered during memory access (i.e., load/store), the BIOS will generate a hardware interrupt to notify the OS that a memory UCE has happened. To deal with these paging errors, including swap-in errors and UCEs, the common wisdom is to terminate the related process or even restart the whole system. Undoubtedly, this ``brute force''  method is simple and effective, but also lower the system's availability.

Fortunately, some applications, such as in-memory caching system, can tolerate such memory data corruption/loss as they can recovery data from disk or replicas. Therefore, in Lightswap, we propose an error handling framework, which provides an opportunity for applications to tolerate and correct these errors in the application context. When a paging error happens, the corresponding application is notified and then the application will try handling this error using its specific error handling routine. 
\section{\emph{Lightswap} Design}
\label{sec:4}

In this section, we first introduce the overview of Lightswap framework and its building blocks. Then we discuss how to effectively handle page fault in user-space and the co-design of swapping and LWT. Finally, we show how paging errors are handled in Lightswap.

\subsection{Lightswap Overview}

\begin{figure*}[t]
  \centering
  \includegraphics[width=0.92\textwidth]{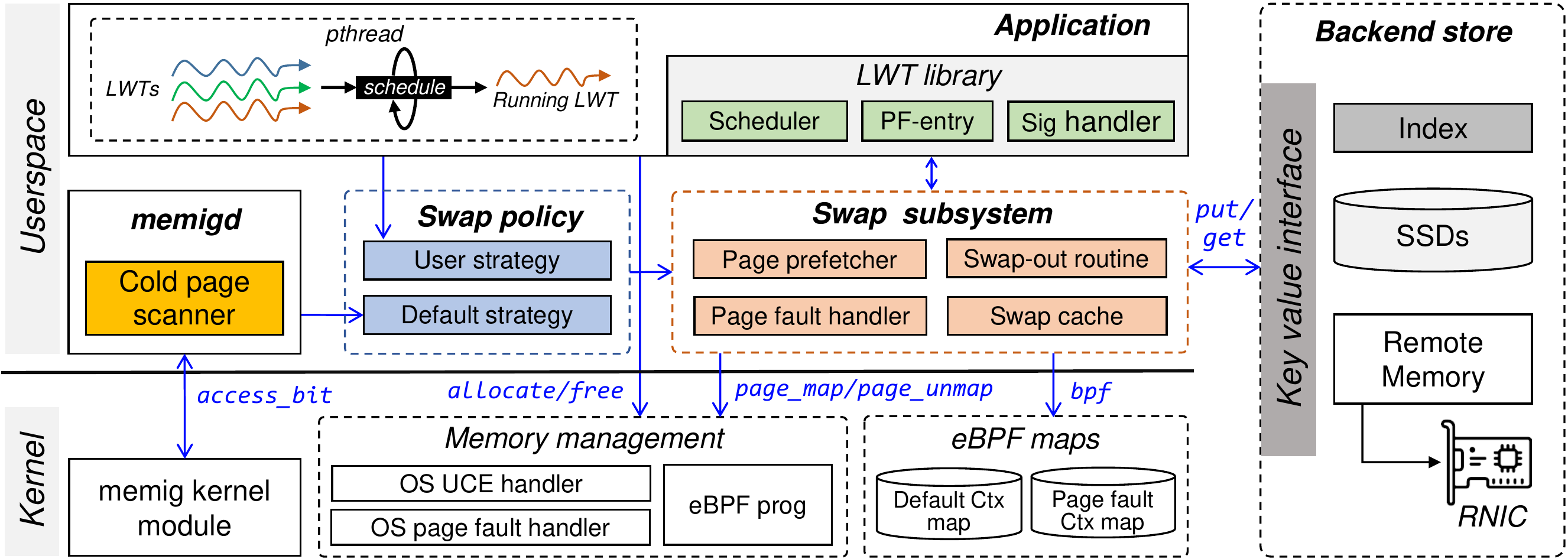} \\
  \caption{Lightswap architecture. Lightswap handles page faults in user-space and swap in/out pages uses a key-value interface.}
  \label{fig: arch}
  \vspace{-0.1in}
\end{figure*}

Figure~\ref{fig: arch} illustrates the overall architecture of Lightswap. As shown, Lightswap handles page faults in user-space and uses a generic key-value store for swapping in/out. For the key-value store, keys are the process virtual addresses, while values are pages. Using key-value interface for swapping makes pages can be swapped to arbitrary storage devices. With one-side RDMA semantics, memory disaggregation can also be enabled by swapping pages to remote memory pools. The components of Lightswap are introduced as below.

\noindent \textit{\textbf{LWT library.}} An application can create multiple standard pthreads, which usually are bounded to given CPU cores. The number of pthreads is limited by the number of available CPU cores to minimize scheduling overhead. Inside each pthread, LWTs are created to process user or client requests. LWT library is provided to user application for creating and managing LWTs. In the LWT library, a scheduler is designed for scheduling LWTs based on the priority. Different from thread scheduling and context-switching, which requires kernel involvement, the scheduling of LWTs is fully controlled by LWT library in user-space without any kernel efforts. Therefore, the scheduling overhead of LWT is minimized. In our measurement, LWT switching latency is usually less than 1 microsecond, which is orders of magnitude faster than thread switching (several microsecond).

To handle page faults triggered by LWTs, each pthread has a page fault entry point (\texttt{PF-entry}), which is the entry point of user-space page fault handler. When a page fault happens, this entry point will be reached and then the user-space page fault handler will be involved after reading the page fault information from eBPF maps (\cref{sec4-2}).

The signal handler (\texttt{Sig handler}) is used to receive paging error signals. For memory UCEs, the BIOS will notify the OS to handle the error, and the OS UCE handler will first tries to isolate the faulted memory address and then issues a signal to the user application, more specifically, to the \texttt{Sig handler}. For swap-in errors, signals will be generated and sent by the swap-in LWTs. In the signal handler, application-specific paging error handling routine will be executed to try to resolve the error (\cref{sec4-4}).

\noindent \textit{\textbf{Memory migration daemon (memigd).}} \texttt{memigd} is a user-space process that responsible for scanning cold pages of given applications. The results will be used as the input of default strategy in swap policy to guide the page reclaiming. To identify cold pages accurately, \texttt{memigd} utilizes a kernel module, namely \texttt{memig} to periodically test and clear the access bit of page table entries. The access bit of a page table entry is set to `1' by hardware once the page is touched. Therefore, in \texttt{memigd}, pages whose access bits are survived during two consecutive scans are considered as hot pages, otherwise, they will be regarded as cold pages. When the system's available memory below a pre-defined threshold, \texttt{memigd} will start to scan cold pages of user applications that labeled as swappable, then these cold pages will be selected as victim pages for swapping out. In Lightswap, to reduce the scanning cost and make mission-critical applications fully reside in memory, only applications that marked as swappable will be scanned for paging out. Moreover, to reduce the I/O operations, we also only write dirty victim pages back to backend store, clean victim pages are discarded directly as they already existed in the backend store.

Besides cold page identification, \texttt{memigd} can also be employed to control the physical memory usage of each process, enabling memory quota for user applications. Once the system's memory is under pressure or applications' memory usage exceeds the quota, \texttt{memigd} will starts scan cold pages and notify the \texttt{uswplib} (see below) for swapping out.

\noindent \textit{\textbf{User-space swap library (lswaplib).}} \texttt{uswplib} is the core component of Lightswap. It is a library that enables user-space swapping for applications. \texttt{uswplib} is comprised of two parts: swap policy and swap subsystem. The swap policy decides which pages to be swap out based on both user specific strategy and default strategy. Besides, swap policy can also be used to guide the page pre-fetching when page fault happens. More specifically, applications can pass application-specific strategy to swap policy by calling:

\vspace{0.03in}
\centerline{ {\small \texttt{void swap\_advise(void* addr, size\_t len, bool out).} } }
\vspace{0.03in}

\noindent If parameter \texttt{out} is true, this function provides swap out suggestions. Pages in \textit{(addr, len)} of that application will be preferred for swapping out. In our current implementation, if application memory quota is configured, pages belong to \textit{(addr, len)} will be swapped out once the application's memory quota is full, otherwise, the swap policy will evenly select victim pages (including cold pages recognized by \texttt{memigd} and pages suggested using \texttt{swap\_advice()}) across applications for swapping out. 

If parameter \textit{out} is false, this function provides prefetch hints. Since we do not know when these pages will be used, to reduce the memory footprints, we neither bring in pages in \textit{(addr, len)} immediately nor bring these pages in a consequent page fault handling process. Instead, we bring in pages belong to \textit{(addr, len)} only when an address in this range causes a page fault. In the corresponding page fault handling routine, we prefetch these pages from the backend store. 

The swap subsystem do the actual paging in and out. A dedicated swap-out thread is launched to receive victim pages from swap policy and write them to backend store. When swapping pages out, pages will be first removed from application's page table use \texttt{page\_unmap()} system call and then added to the swap cache. Pages in swap cache are asynchronously writeback to the backend store, thus decoupling page write from the critical path of swapping out routine. A dedicate thread periodically flushes the swap cache to the backend store when its size has reached a pre-configured threshold batch size. \texttt{uswplib} defines an user-space page fault handler, which will be called to swap in the requested page when page fault happens. The page fault handler will first search the swap cache for the desired page. If the page is found in the swap cache, we remove it from the swap cache and add it to the application's page table at the faulted address. Otherwise, page will be read from the bankend store. In the page fault handler, we also decouple the prefetching from the critical path of swapping in routine. After bringing the desired page into memory, the page fault handler appends a prefetch request into a prefetch queue. If the faulted address associated with an application-specific prefetch hint, the prefetch request will read pages in \textit{(addr, len)} specified by \texttt{swap\_advice()}. Otherwise, a simple read-ahead policy will be used and the prefetch request tries to read the surrounding eight pages at the faulted address. To improve the concurrency and of page prefetching, a group of I/O LWTs, referred as \textit{prefetchers} will constantly pull requests from the prefetch queue and bring pages into memory. Note that in our current implementation, we do not have any special or carefully designed prefetch algorithm as this work does not aim at prefetching, but these algorithm can be easily added to Lightswap.

\begin{figure}[t]
	\centering
	\includegraphics[width=0.9\columnwidth]{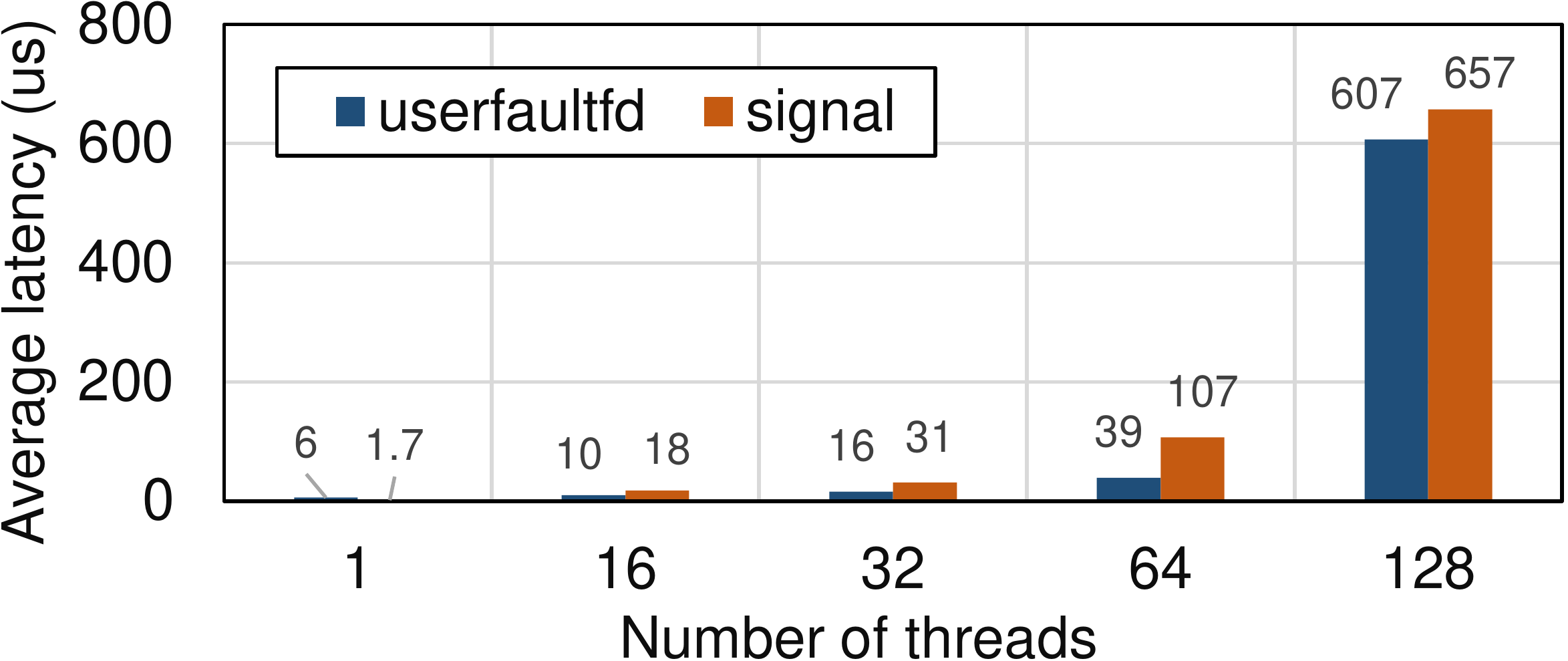} \\
	\caption{Average page fault events notification latency under different number of threads.}
	\label{fig: pf-latency}
	\vspace{-0.15in}
\end{figure}

\noindent \textit{\textbf{eBPF prog.}} The \texttt{eBPF prog} is eBPF bytecode that injected into the kernel using \texttt{bpf()} system call. It is the key component of handling page fault in user-space (\cref{sec4-2}). eBPF prog maintains two context maps: default context map and page fault context map. Both maps contains multiple entries. Each entry of the default context map stores the default context that is saved at page fault entry point (i.e., \texttt{PF-entry}). Each entry of the page fault context map stores the LWT context where page fault is triggered. When page fault happens, \texttt{eBPF prog} is responsible for 1) saving the context of faulting LWT into the page fault context map and 2) modify the current context with the previously saved default context.

\subsection{Handling Page Fault in User-space}
\label{sec4-2}

\begin{figure}[t]
	\centering
	\includegraphics[width=0.9\columnwidth]{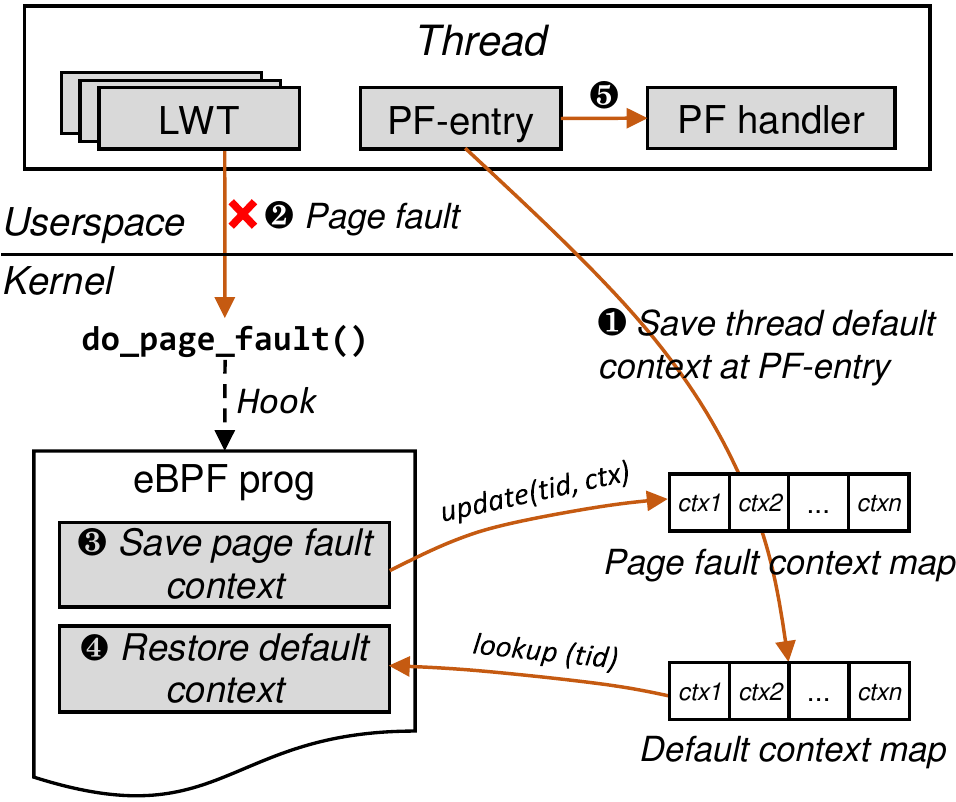} \\
	\caption{eBPF-based page fault handling scheme.}
	\label{fig: ebpf-fault}
	\vspace{-0.1in}
\end{figure}

Page faults are handled in user-space in Lightswap. The key challenge here is how to notify user-space application when page fault happens effectively. To achieve this, Linux provides userfaultfd~\cite{userfault} to notify and allow user-space applications handle page faults of pre-registered virtual address regions. Besides, one can also use Linux signal to notify user-space applications that page faults happened, which is similar to the memory UCE notification (\cref{sec4-4}). However, both userfaultfd and signal suffer from several performance issues as discussed below.

Userfaultfd requires a fault-handling thread that pulls page fault events through reading the userfaultfd file descriptor, and provides \texttt{UFFD\_COPY} and \texttt{UFFD\_ZERO} ioctl operations to resolve the page faults. When a page fault happens, The OS page fault handler puts the faulting thread into sleep and allocates physical page for the faulted address, then an event is generated and sent to the fault-handling thread. The fault-handling thread reads events from userfaultfd file descriptor and resolves the page fault with \texttt{UFFD\_COPY} or \texttt{UFFD\_ZERO} ioctl operations. In userfaultfd, all the page faults are handled by the fault-handling thread, which will easily become the bottleneck when multiple threads trigger page faults almost simultaneously. Moreover, one cannot directly bring the request page from backend store to the faulted address. To resolve a page fault, one must first read the request page from backend store to a memory buffer, and then copy the page from the buffer to the faulted address using \texttt{ioctl} system call, which brings one extra memory copy operation.

For the signal approach, when page faults happen, the OS page fault handler notifies the user-space by sending a signal, which contains the faulted address and other related information. However, as the signal handler of a process is shared by all its threads, a lock is required to protect the signal handler data structure, which makes the signal sending routine suffers from seriously lock contention under high-concurrency.

To show the performance of userfaultfd and signal, we record the page fault notification latency (i.e., latency from page fault happens to the user-space page fault handler receives the page fault event) of both userfualtfd and signal under different concurrent threads and plot their average latency in Figure~\ref{fig: pf-latency}. The detailed configurations can be found in \cref{subsec: microbench}. As shown, with only one thread, both userfaultfd and signal achieve very low page fault notification latency (i.e., 6$\mu$s for userfaultfd and 1.7$\mu$s for signal). However, with the increasing of concurrent threads, the page fault latency of both userfaultfd and signal increase significantly. Therefore, neither userfaultfd nor signal is impractical for handling page fault in user-space for high-currency applications.

\begin{figure}[t]
	\centering
	\includegraphics[width=0.9\columnwidth]{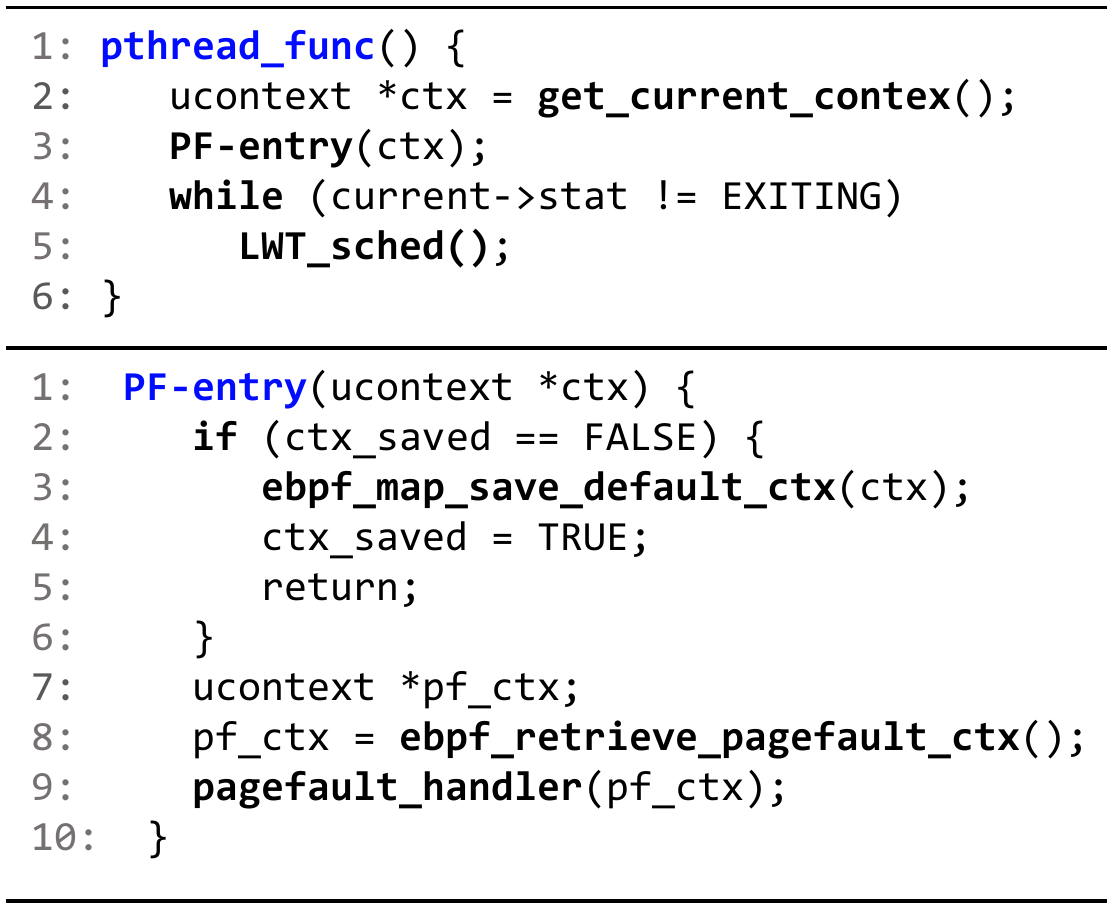} \\
	\caption{Swapping with LWT. Thread schedules LWTs and calls page fault handler when any LWT triggers page fault.}
	\label{fig: lwt-sched}
	\vspace{-0.1in}
\end{figure}

To effectively handling page faults in user-space, we propose the eBPF-based page fault notification scheme. As shown in Figure~\ref{fig: ebpf-fault}. In the main thread, before launching and scheduling LWTs, thread will enter the page fault entry point (i.e., \texttt{PF-entry}) and saves the current thread context into the default context map (\ding{182}). When one of the LWT triggers a page fault (\ding{183}), the kernel page fault handler will be involved. We use the kernel page fault handling function (i.e., \texttt{do\_page\_fault()}) as a hook point and attached our eBPF program to this function. Once this function is involved, the attached eBPF program will be executed. In the eBPF program, we first save LWT's context at the point that page fault happens (\ding{184}). We refer this context as page fault context and store it into the page fault context map, which uses the thread ID (i.e., \texttt{tid}) as the key and the thread context as its value. The page fault context contains the page fault address and will be used to restore the execution of LWT after page fault is resolved. Then the thread's default context (which is saved in step \ding{182}) is retrieved from the default context map, and the current thread context is modified to the retrieved default context, which makes the execution of current thread restore to \texttt{PF-entry} (\ding{185}). In \texttt{PF-entry}, thread will notice that the default context is already saved, which means this is not the first enter and thus thread knows page fault occurs. Then, the page fault context is also retrieved from the page fault context map and saved in the faulting LWT's stack. The page fault context will be employed by the LWT scheduler to restore the execution of the faulting LWT. Finally, the user-space page fault handler is called to resolve the page fault (\ding{186}). In the user-space page fault handler, the faulting LWT is blocked and put into sleep, then the requested page will be read from the backend store (see details in \cref{sec4-3}). Once the page fault is resolved, the state of the faulting LWT is set to runnable and will be scheduled in the next scheduling period. In the next subsection, we will show how pages are read from backend store using LWT.

\begin{figure}[t]
	\centering
	\includegraphics[width=0.95\columnwidth]{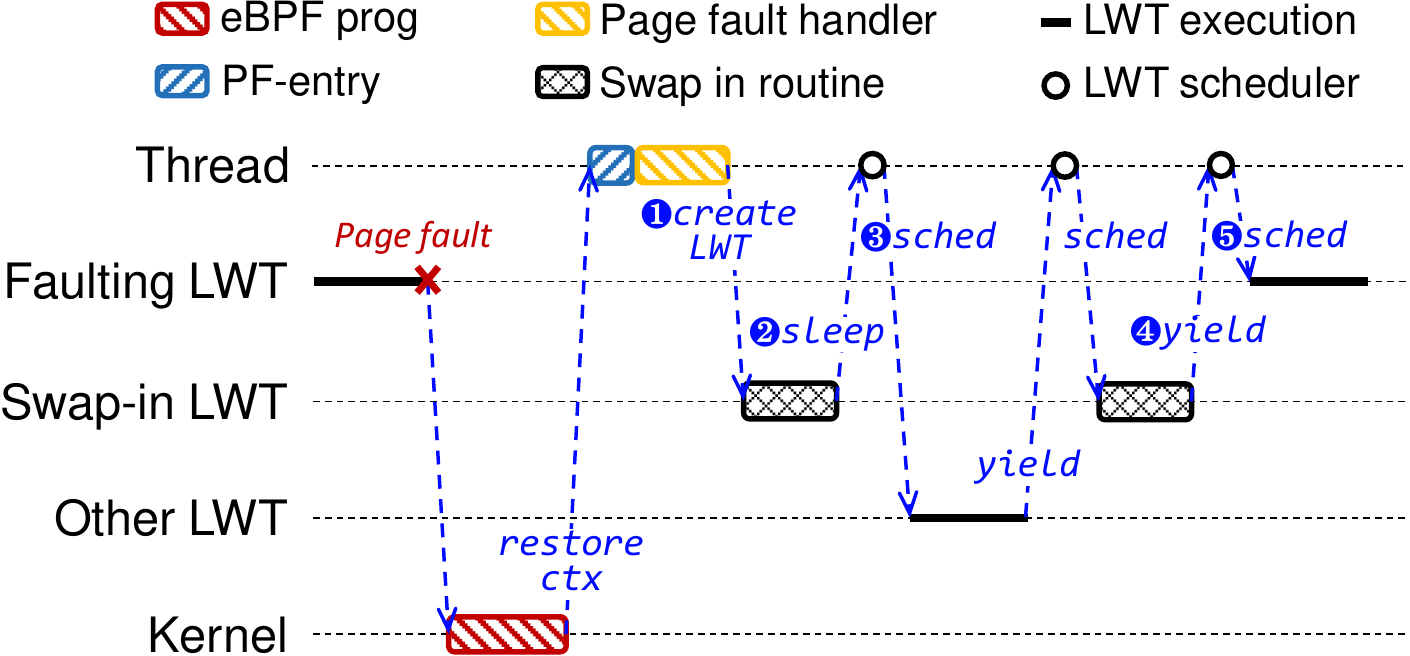}	
	\caption{Scheduling of swap-in LWTs and faulting LWTs.}
	\label{fig: swapin-sched}
	\vspace{-0.15in}
\end{figure}


\subsection{Co-design Swapping with LWT}
\label{sec4-3}

Lightswap handles page faults in user-space. There are two challenges that we need to address. First, user applications must use Lightswap transparently to avoid application modifications. Second, to reduce the total page fault latency, the faulting LWT must be woken up as soon as possible after the requested page is brought into memory.


To address these issues, we co-design swapping with LWT to reduce the swap-in latency. Figure~\ref{fig: lwt-sched} shows the pseudo code of how LWTs are scheduled and user-space page fault handler are called. In the main thread, the thread's context is saved to the default context eBPF map in \texttt{PF-entry()} before scheduling LWT. After that, the LWT scheduler (i.e., \texttt{LWT\_sched()}) continuously picks and run LWTs from the front of the ready LWT queue. When any LWT triggers page fault, the faulting LWT is blocked and the thread is restore to \texttt{PF-entry()}, in which the user-space page fault handler will be called. To tackle the first challenge, a new LWT (referred as \textit{swap-in LWT}) is created to swap in the requested page in the page fault handler. Thus, user application does not aware page fault happens and the page fault will be handled by our dedicated swap-in LWT. To tackle the second challenge, we make the LWT scheduler prefers swap-in LWTs and faulting LWTs. To achieve this, we classify the ready LWTs into three queues: 1) \textit{swap-in LWT queue}, swap-in LWTs are put into this queue after they are created and ready to run; 2) \textit{faulting LWT queue}, which contains LWTs that encounter page faults and the page faults have been resolved, which means their requested pages have been brought into memory by the swap-in LWTs and their status become ready; 3) \textit{normal LWT queue}, other ready LWTs are resided in this queue. The LWT scheduler assigns the first priority to the swap-in LWT queue, second priority to the faulting LWT queue, and third priority to the normal LWT queue. Therefore, swap-in LWTs can be scheduled to run immediately after they are created, and once the requested pages are swapped back to memory, the faulting LWTs can be scheduled to run as soon as possible.

\begin{figure}[t]
	\centering
	\includegraphics[width=0.9\columnwidth]{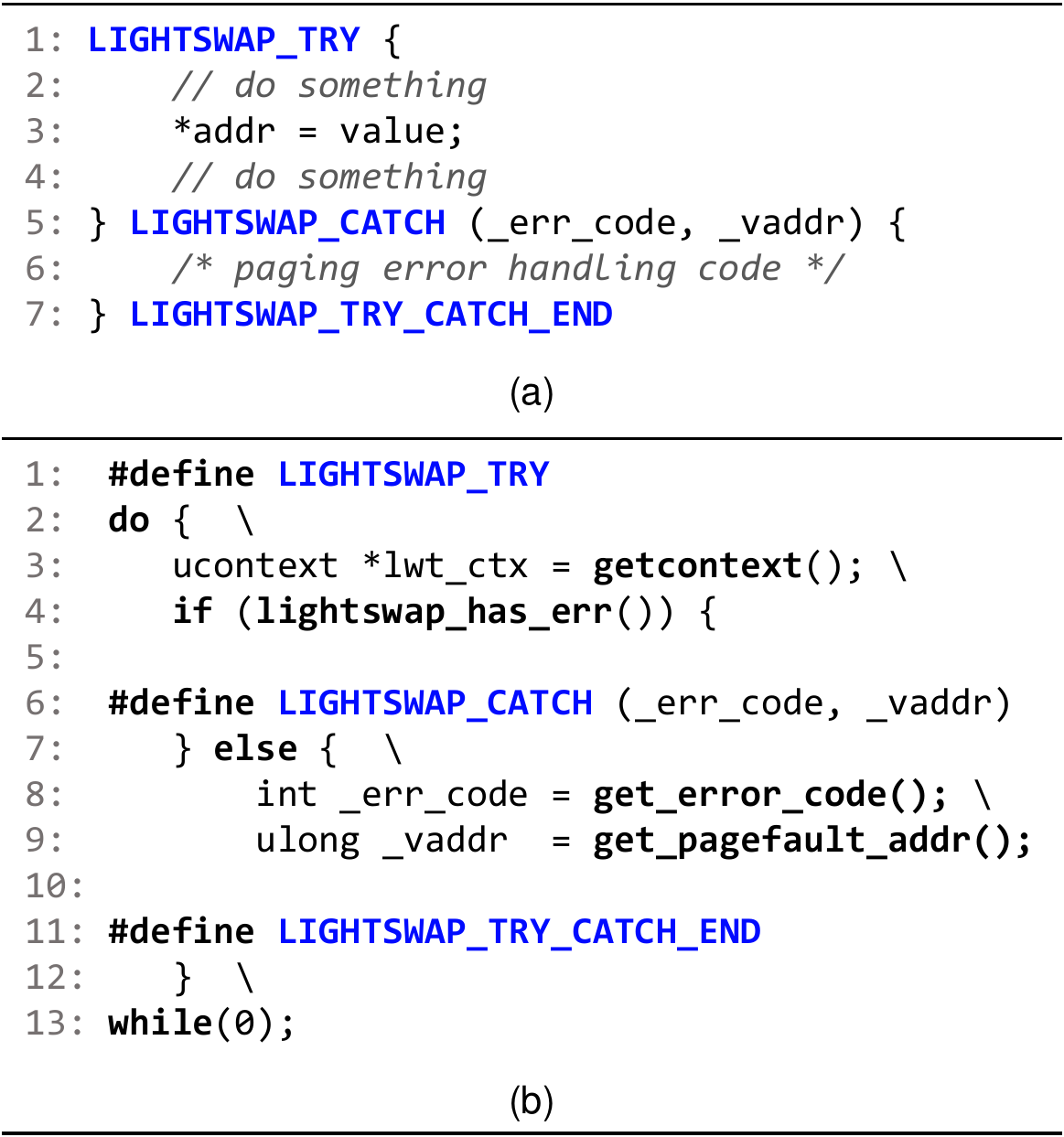} \\
	\caption{Try-catch paging error handling framework. \textit{(a) Example of how application handling paging errors with the proposed try-catch framework; (b) Lightswap try-catch keywords macro definition.}}
	\label{fig: try-catch}
	\vspace{-0.1in}
\end{figure}

Figure~\ref{fig: swapin-sched} illustrates an example of scheduling of swap-in LWT that reading page from backend store. As shown, when page fault happens, the faulting LWT is blocked and a dedicated swap-in LWT is created in the user-space page fault handler (\ding{182}). We assume that the swap-in LWT queue is empty currently, thus when the swap-in LWT is added to the swap-in LWT queue, it will be scheduled to run immediately. After the requested page is brought into memory, the swap-in LWT changes the state of faulting LWT to ready and adds it to the faulting LWT queue. Then it gives up the CPU by calling \texttt{yield()}  (\ding{185}). Finally, the scheduler picks the faulting LWT from the queue and schedules it to run (\ding{186}). To improve the throughput, we propose swap in pages asynchronously by leveraging the negligible scheduling overhead of LWT. As shown in Figure~\ref{fig: swapin-sched}, the swap-in LWT is put into sleep when waiting for page to be read from backend store (\ding{183}). Thus, other ready LWT can be scheduled to run to maximize the CPU usage(\ding{184}). After the page is brought into memory, the swap-in LWT is set to ready and re-added to the swap-in LWT queue. Remember that swap-in LWT queue has the highest priority, and thus the swap-in LWTs will be preferred by the scheduler at the next LWT scheduling.

\subsection{Try-catch Exception Framework}
\label{sec4-4}

\begin{figure}[t]
	\centering
	\includegraphics[width=0.92\columnwidth]{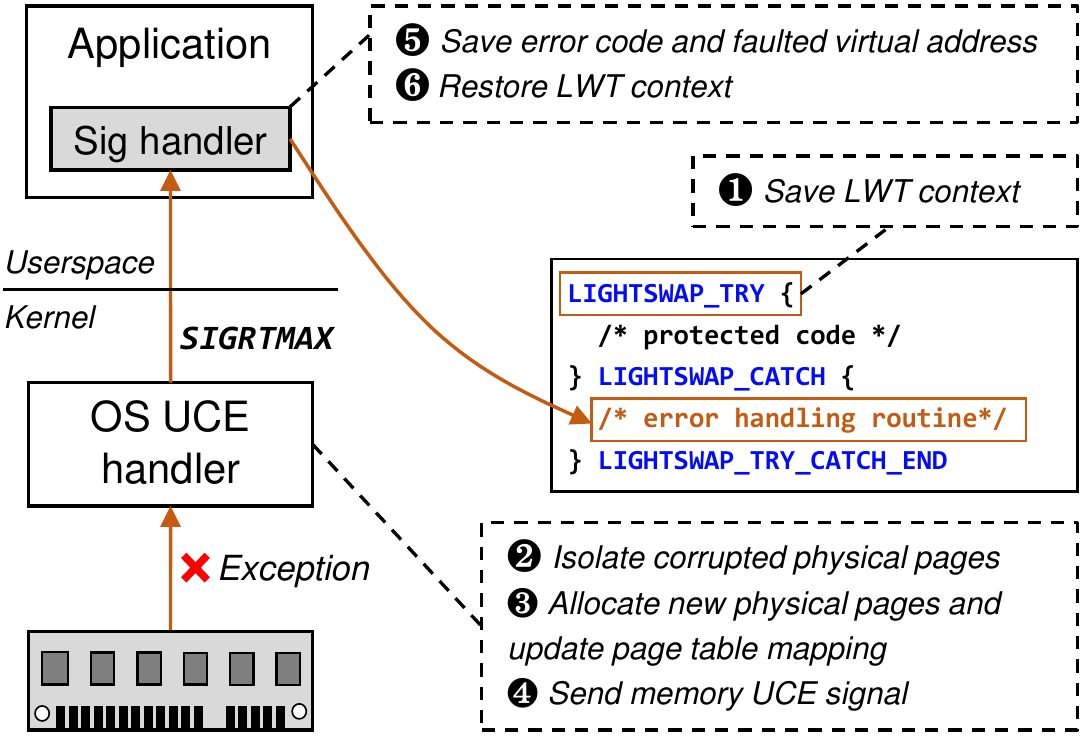} \\
	\caption{Handling memory UCE.}
	\label{fig: handle-uce}
	\vspace{-0.15in}
\end{figure}

Inspired by the try-catch exception handling approach in C++, we designed a paging error handling framework in Lightswap. Basically, applications can embed \texttt{LIGHTSWAP\_TRY} and \texttt{LIGHTSWAP\_CATCH} macro into their program, as shown in Figure~\ref{fig: try-catch}(a). Codes that surrounded by \texttt{LIGHTSWAP\_TRY} macro will be protected against from paging errors. For the example in Figure~\ref{fig: try-catch}(a), if the pointer deference at line 3 triggers a paging error (memory UCE or swap-in error), the application will jump to the \texttt{LIGHTSWAP\_CATCH} immediately to handle the paging error using application customize codes. For example, memory cache applications can recovery the data from disk. Through this way, we provide an opportunity for applications to handle paging errors in user-space.

Figure~\ref{fig: try-catch}(b) shows the definition of Lightswap try-catch macro. In \texttt{LIGHTSWAP\_TRY} macro (line 1--4), we first save the context of the current LWT (line 3) and check whether a paging error is encountered (line 5). In normal execution, \texttt{lightswap\_has\_err()} returns false and codes in the \texttt{LIGHTSWAP\_TRY} block will be executed. If a paging error happens when executing codes in the \texttt{LIGHTSWAP\_TRY} block, the user-space page fault handler or the signal handler will restore the program pointer to the context saving point (line 3) using the previous saved LWT context. Then, function \texttt{lightswap\_has\_err()} will return true as an error happens, which makes the application jumps to \texttt{LIGHTSWAP\_CATCH} block to handle the error.

Figure~\ref{fig: handle-uce} shows how memory UCE is handled in Lightswap. For memory access that causes memory UCE, the OS UCE handler will first be notified by the hardware. In the OS UCE handler, the corrupted physical pages are isolated and new pages are allocated and mapped to the faulted virtual addresses (\ding{182}\ding{183}). Then a standard Linux signal, which contains the signal number, error type (i.e., memory UCE), and faulted virtual address is sent to the corresponding application (\ding{184}). Currently, we utilize the maximum signal number (i.e., \texttt{SIGRTMAX}) as the memory UCE signal. After the signal handler captures the signal, it first saves the error code and faulted virtual address, which will be used in the \texttt{LIGHTSWAP\_CATCH} block, then it restore LWT context to the point that the context is saved (\ding{185}\ding{186}). With these steps, the application will finally jump to the \texttt{LIGHTSWAP\_CATCH} block to handle the memory UCE.

To handle swap-in errors, the user-space page fault handler is responsible for restoring the LWT context. For example, if the pointer deference at line 3 in Figure~\ref{fig: try-catch}(b) triggers a page fault and the user-space page fault handler finds that the requested page cannot be brought in memory correctly, it first saves the error code and faulted virtual address, and then restores the LWT context to let application to handle the swap-in error.

\section{Evaluation}
\label{sec:5}

This section introduce the evaluation of Lightswap, we start with a brief introduction of our system implementation, then give the evaluation setups and finally discuss the results.

\begin{table}[t]
	\centering
	\footnotesize
	\setlength{\tabcolsep}{0.94em}
	\begin{tabular}{l|c|ccccc}
		\hline
		\multicolumn{2}{c|}{\textbf{Number of threads/LWTs}}  & 1 & 16 & 32 & 64 & 128 \\
		\hline
		\multirow{3}{*}{\bf{Latency ($\mu$s)}} & \textbf{Userfaultfd} & 6 &	10  & 16 & 39 & 607 \\
		& \textbf{Signal}      & 1.7 & 18 & 31 & 107 & 657  \\
		& \textbf{eBPF}        & 1.6	& 1.7 & 2 & 2.4 & 4 \\
		\hline
	\end{tabular}
	\caption{Average latency of different user-space page fault notification schemes.}
	\label{tbl: pf-notify-lat}
	\vspace{-0.3in}
\end{table}

\subsection{Implementations}
We implement and evaluate Lightswap in real production system to show its effectiveness. We made some modifications to the Linux kernel to let it supports user-space swapping effectively. First, we add a hook point, more especially, a empty function to the kernel. The kernel page fault handler (i.e., \texttt{do\_page\_fault()}) will call this function and then return immediately if the faulted address belongs to an application that is supported by Lightswap. We make the eBPF program to hook this function and thus normal page faults still use the kernel page fault handler while only applications that are supported by Lightswap use our user-space page fault handler. To reduce the number of \texttt{bpf()} system calls, we use shared memory between kernel and user-space to share the page fault context map. Second, to support swap in/out pages in user-space, we added a pair of system calls (i.e., \texttt{page\_map()} and \texttt{page\_unmap()}) to respectively map or unmap a page to or from a given virtual address, thus the swap-in LWT can update the page table mapping for the faulted address, and the swap-out thread can also remove a page from the application's page table. Third, we modified the OS UCE handler to make it send memory UCE signal to our signal handler if the OS cannot correct the memory error. Totally, all these kernel modification effort is no more than 1000 lines of code.

To implement the key-value based backend store, we use a in-memory hash table as its index to reduce the index traversal time. We use a second hash table to solve the hash conflicts. Entries in the second hash table point to the actual positions of pages. For local SSDs, we organize the SSD space in a log-structured way and thus pages in swap cache can be flushed in batches to maximize the throughput. For remote memory, we deploy a daemon in remote memory servers to reserve and allocate memory space. To reduce the number of allocation requests, memory servers only allocate 1GB large memory blocks and response clients their registered memory region IDs and offsets for RDMA. The backend store in the client is responsible for managing memory blocks and splitting them into pages.

\subsection{Evaluation Setups}
We employ two x86 servers in the evaluation, one is used as client for running applications. Another server is configured as memory server to allocate memory blocks. Each server equips with two Intel Xeon CPUs, and each CPU contains 40 cores with hyper-thread enabled. The memory capacity of both server is 256GB. We will limit the memory usage of client server in order to trigger swapping in/out. The connection between two serves is 100G RoCE with our customized user-space driver that based on DPDK. For paging with local SSDs, we use the state-of-art NVMe SSD with our SPDK-based user-space NVMe driver as the storage device.

\begin{figure}[t]
	\centering
	\includegraphics[width=0.95\columnwidth]{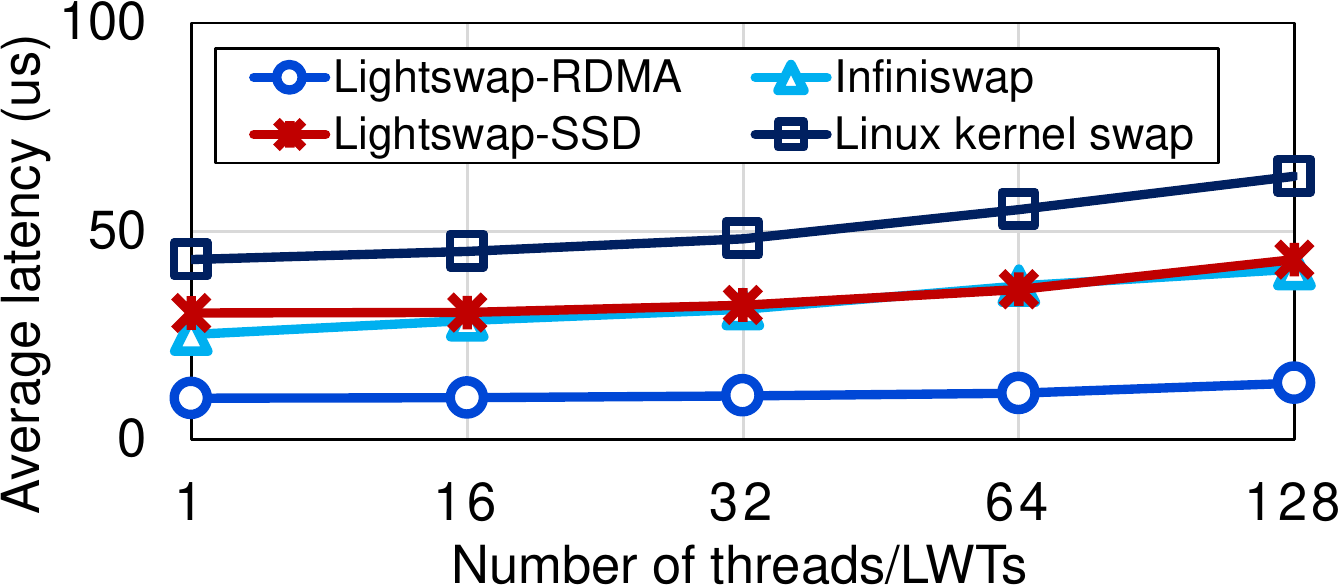} \\
	\caption{Average page fault handling latency, from page fault happens to the requested page be brought into memory by the page fault handler. \textit{Lightswap-RDMA denotes paging with remote memory via one-side RDMA, while Lightswap-SSD represents paging with local SSDs.}}
	\label{fig: pf-handle-latency}
	\vspace{-0.2in}
\end{figure}

\begin{figure*}[t]
	\centering
	\includegraphics[width=0.98\textwidth]{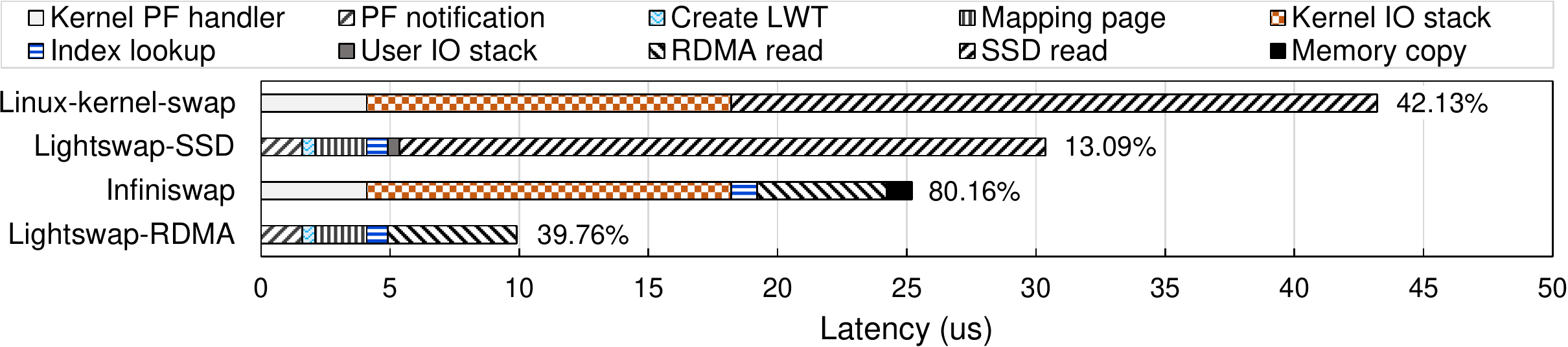} \\
	\caption{Page fault handling latency breakdown under no concurrency. \textit{The numbers beside each bar denote the fraction of time cost by the software stack during handling page faults.}}
	\label{fig: lat-breakdown}
	\vspace{-0.1in}
\end{figure*}

\subsection{Microbenchmarks}
\label{subsec: microbench}

\textbf{Page fault notification latency.} To show the effectiveness of eBPF-based user-space page fault handling, we first evaluate the average page fault notification latency under different concurrency, where the page fault notification latency denotes the latency from page fault happens to the user-space page fault handler receives the page fault event. We compares the average page fault notification latency among eBPF, userfaultfd and signal. Since our eBPF-based approach is co-designed with LWT, to evaluate its performance, we create multiple threads (from 1 to 128) and bound them to certain CPU cores. Inside each thread, we launch a LWT as the faulting LWT. For userfaultfd, we use one faulting-handling thread to pull page fault event and create multiple threads as faulting threads. For the signal approach, a signal handler is registered as the user-space page fault handler. In this scheme, we reuse the signal number of handling swap-in errors and memory UCEs, and use error code to identify the actual fault type (i.e., page fault, swap-in error or memory UCE). In the page fault handler of all these schemes, we simply allocate and map a zeroed page for the faulted address and return immediately.

Table~\ref{tbl: pf-notify-lat} shows the average page fault notification latency of handling page faults in user-space. As shown, when there is no concurrency, all of these page fault notification schemes perform well, achieving extreme low latency. However, with the increase of concurrency, the notification latency of both userfaultfd and signal increase exponentially. With 128 threads, the average latency of userfaultfd and signal as high as 607$\mu$s and 657$\mu$s, respectively. As we discussed in \cref{sec4-2}, for userfaultfd, the high latency is caused by the contention of fault handling thread. The fault handling thread can launch multiple threads to handle page faults concurrently, but this also brings extra CPU overheads and adds synchronous costs. For the signal scheme, in the signal sending routine (i.e., \texttt{force\_sig\_info()}), a \texttt{siglock} must be obtained before sending the signal, which leads to seriously lock contention under high-concurrency and thus resulting the high latency of page fault notification. In the contrary, the proposed eBPF-based page fault notification scheme achieve extreme low latency under all the degrees of concurrency. When the number of LWTs increases from 1 to 128, the average latency only has a slight increment, which is mainly due to the lock contention of eBPF maps. In our current implementation, we employ 32 eBPF hash maps for page fault context, and a lock is used to protect each map. We divide the page fault contexts of LWTs into these eBPF maps evenly by using the core ID as an index number. Thus, even with 128 LWTs, each eBPF map only needs to store the page fault contexts of 4 LWTs, which significantly reduces the lock contention and contributes to the slight increment of latency under high-concurrency.

\begin{figure*}[t]
	\footnotesize
	\begin{minipage}{0.984\textwidth}
		\centerline{\includegraphics[width=1\textwidth]{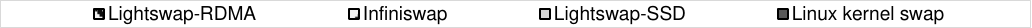}}
	\end{minipage}
	
	\begin{minipage}{0.32\textwidth}
		\centerline{\includegraphics[width=1\textwidth]{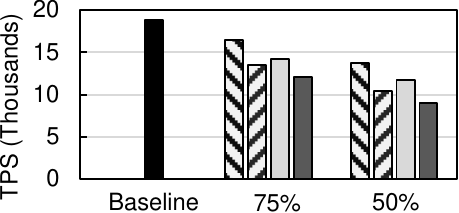}}
		\centerline{(a) Readmost}
	\end{minipage}
	\enspace
	\begin{minipage}{0.32\textwidth}
		\centerline{\includegraphics[width=1\textwidth]{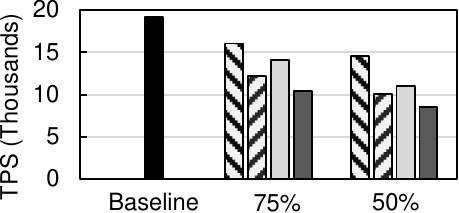}}
		\centerline{(b) Readwrite}
	\end{minipage}
	\enspace
	\begin{minipage}{0.32\textwidth}
		\centerline{\includegraphics[width=1\textwidth]{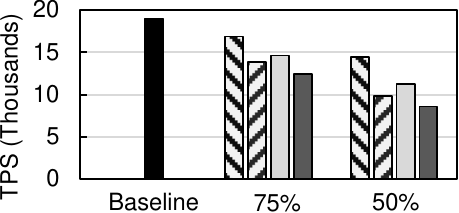}}
		\centerline{(c) Writemost}
	\end{minipage}
	
	\caption{Average TPS of memcached with different swapping schemes. \textit{We compare the performance of different swapping schemes with 75\% and 50\% physical memory of the memcached dataset size. We use 32 worker threads to process requests for all these configurations.}}
	\label{fig: cache-tps}
	\vspace{-0.1in}
\end{figure*}

\noindent \textbf{Page fault handling latency.} To show the end-to-end performance of Lightswap, we compare the page fault handling latency of Lightswap to other swapping schemes. We denotes the page handling latency as the time duration from page fault happens to the page fault handler finishes resolving the page fault. We compare the results between Lightswap, Infiniswap and the Linux default kernel swap. Figure~\ref{fig: pf-handle-latency} illustrates our evaluation results. Note that page fault handling latency does not include the time duration from the page fault be resolved to the point that the faulting thread/LWT is restored to run. As shown, when paging with remote memory through one-side RDMA, Lightswap achieves the lowest page fault handling latency, ranges from 10$\mu$s to 13.5$\mu$s under different degrees of concurrency. The conventional Linux kernel swap has the highest page fault handling latency, ranges from 43.2$\mu$s to 63.2$\mu$s. When paging with remote memory via one-side RDMA, Lightswap respectively outperforms Infiniswap and Linux kernel swap by around 2.5 - 3.0 times and 4.3 - 5.0 times in terms of page fault handling latency. Even paging with local SSDs, the proposed Lightswap achieves comparable performance with Infiniswap, and has around 30\% lower latency than Linux kernel swap. With ultra-low latency NVMe SSDs, such as Intel Optane and KIOXIA XL-Flash, we believe that Lightswap-SSD can achieve lower latency than Infiniswap.

To demystify the reason behind this improvements, we breakdown the page fault handling process and plot its detailed time cost in Figure~\ref{fig: lat-breakdown}. In the figure, the kernel PF (page fault) handler has already included the time spend on trap into the kernel. In our measurement, reading a 4KB page from remote memory via one-side RDMA and local SSD will respectively cost 5$\mu$s and 25$\mu$s on average in our environment. Lightswap handles page faults in user-space and avoids the slow kernel data path by leverage high-performance user-space drivers. Therefore, in Lightswap, the page fault handling latency is dominated by the page read latency. Software stack respectively takes 39.76\% and 13.09\% for paging with remote memory and local SSDs. However, both Infiniswap and Linux kernel swap need to go through the entire kernel I/O stack when fetch pages, making the kernel I/O stack takes a large fraction of the total latency. The kernel I/O stack is mainly comprised by the generic block layer that provides OS-level block interface and I/O scheduling, and the device driver that handles device specific I/O command submission and completion. As shown in the figure, due to the kernel I/O stack, the software stack overheads for Infiniswap and Linux kernel swap are 80.16\% and 42.13\%, respectively. Despite the fact that Infiniswap also one-side RDMA, the high software stack overhead makes its page fault handling latency reaches 25$\mu$s, and even exceeds 40$\mu$s under high-concurrency.

\subsection{Application: Memcached}

Memcached is an widely used in-memory key-value based object caching system. Memcached uses the client-server mode and in the server sides, multiple worker threads is created to process the PUT and GET requests from the client side. We benchmark memcached with the YCSB workloads~\cite{yscb} under different swapping schemes. Each YCSB workload performs 1 million operations on memcached with 10,485,760 1KB records, which are 10GB data in total. Table~\ref{tbl: workloads} summarizes the characteristics of our YSCB workloads.

\begin{table}[h]
	\centering
	\footnotesize
	\setlength{\tabcolsep}{0.68em}
	\begin{tabular}{c|cccccc}
		\hline
		\bf{Workload Name} & \bf{Read} & \bf{Insert} & \bf{Update} & \bf{OPs} & \bf{Size} \\
		\hline
		Readmost  & 90\% & 5\%  & 5\%  & 1 million & 10GB \\
		Readwrite & 50\% & 25\% & 25\% & 1 million & 10GB \\
		Writemost & 90\% & 5\%  & 5\%  & 1 million & 10GB \\
		\hline
	\end{tabular}
	\caption{YCSB workloads characteristics.}
	\label{tbl: workloads}
	\vspace{-0.25in}
\end{table}

\begin{figure}[t]
	\footnotesize
	\begin{minipage}{0.995\columnwidth}
		\centerline{\includegraphics[width=1\textwidth]{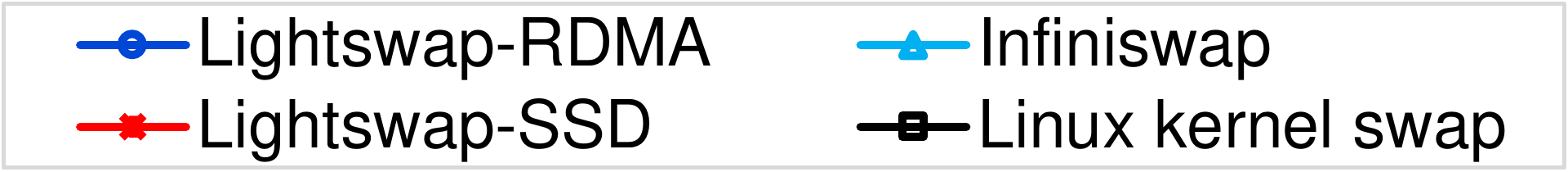}}
	\end{minipage}

	\begin{minipage}{0.315\columnwidth}
		\centerline{\includegraphics[width=1\textwidth]{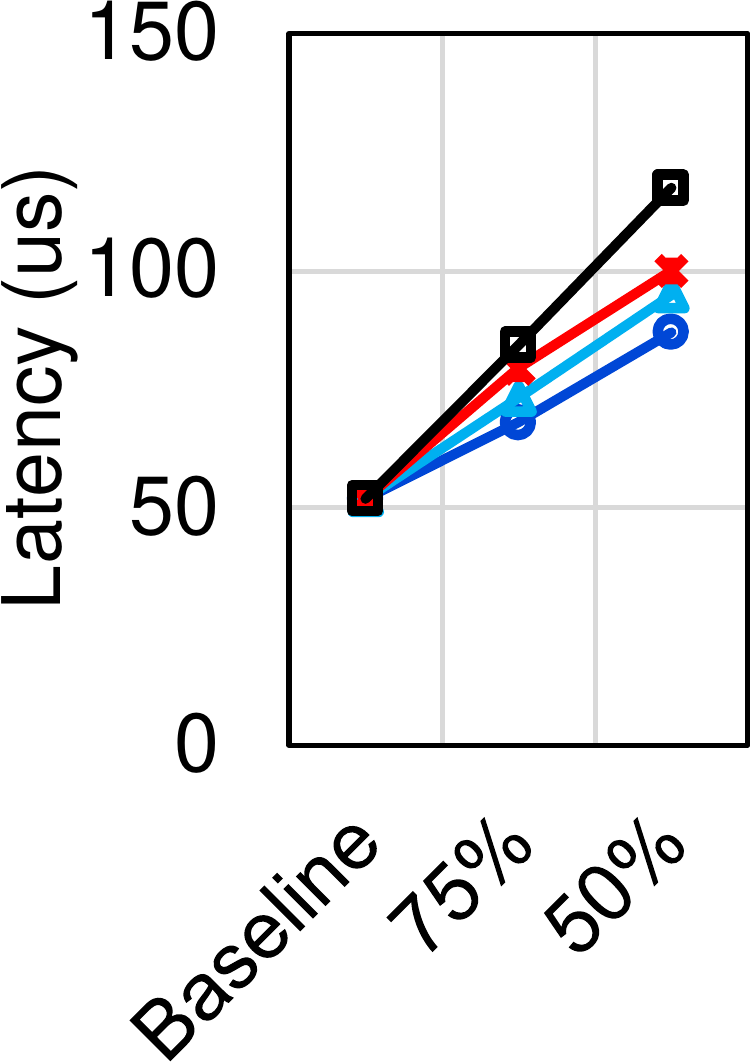}}
		\centerline{(a) Readmost}
	\end{minipage}
	\enspace
	\begin{minipage}{0.315\columnwidth}
		\centerline{\includegraphics[width=1\textwidth]{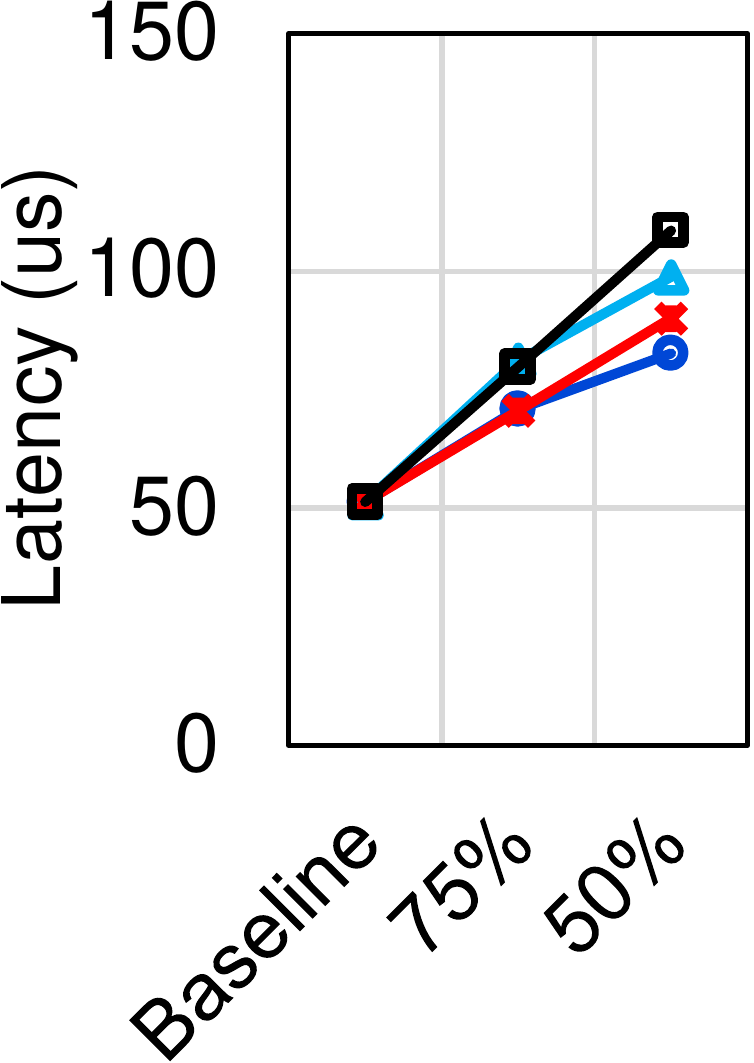}}
		\centerline{(b) Readwrite}
	\end{minipage}
	\enspace
	\begin{minipage}{0.315\columnwidth}
		\centerline{\includegraphics[width=1\textwidth]{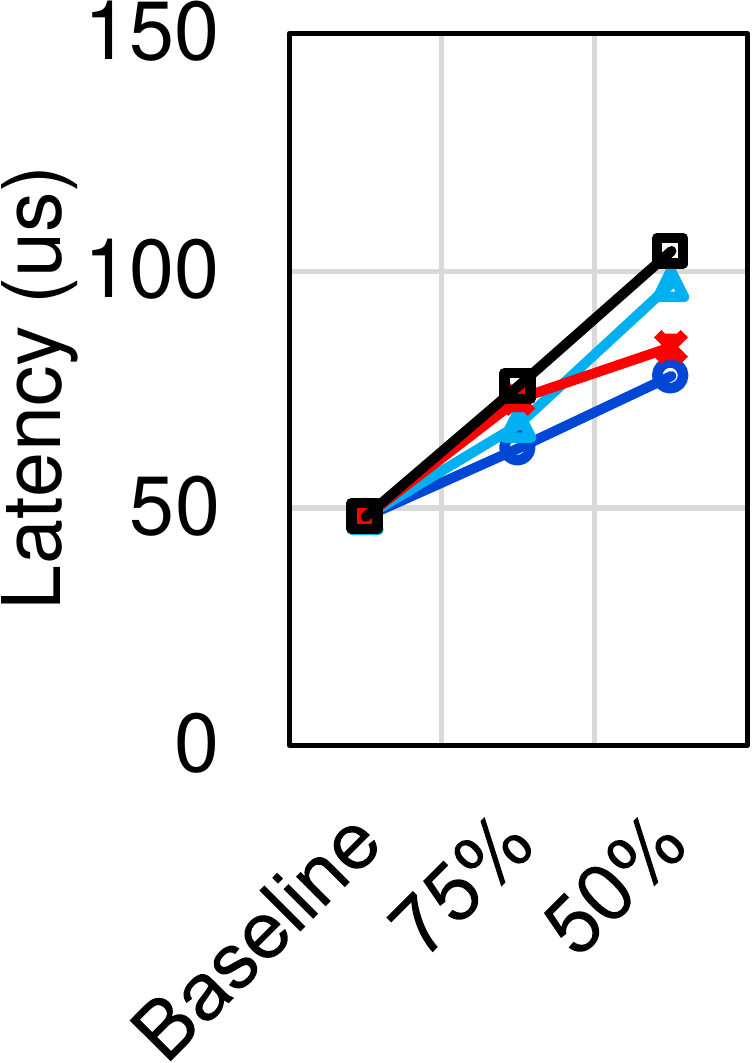}}
		\centerline{(c) Writemost}
	\end{minipage}
	
	\caption{Comparison of average operation latency under different swapping schemes.}
	\label{fig: cache-lat}
	\vspace{-0.2in}
\end{figure}

In memcached, mutiple worker threads are created to handle requests from the client-side. However, the proposed Lightswap is co-designed with LWT, so we first rewrite memcached and make it to use LWTs to process client-side requests. In this LWT version memcached, we create mutiple worker threads and bound these threads to certain CPU cores. Inside each thread, we launch a worker LWT for each incoming request. We limit the number of LWTs in each thread to 10 to reduce the LWT management overhead. Since the LWT execution mode is much similar to the thread mode, the rewritting does not need too much efforts. In our system's LWT implementation, each LWT has a maximum 32KB stack, and the execution of LWTs is non-preemptive, which means they will hold the CPU till get terminated or reach a waiting state (e.g., waiting semaphore) or proactively release the CPU by calling \texttt{yield()}. For other swapping schemes, we still use the thread version memcached.

Figure~\ref{fig: cache-tps} compares the average throughput of different swapping schemes. Besides, we also compares their average operation latency in Figure~\ref{fig: cache-lat}. In these two figures, The baseline indicates the case that all memcached's dataset is reside in memory and there is no swap-ins/-outs. Since the number of threads/LWTs in memcached is smaller than the number CPU cores (i.e., 80 in total), all the worker threads/LWTs can occupy the CPU continuously. Therefore, we observed that there is no performance difference between LWT version and thread version memcached. We found 3 key results from these figures: 1) Lightswap-RMDA has the least performance degradation, it outperforms Infiniswap and Linux kernel swap by 40\% and 60\% on average in throughput, respectively; 2) Due to the outstanding page fault handling latency, Lightswap-RDMA also achieves the lowest latency among these swapping schemes, it outperforms Infiniswap and Linux kernel swap by 18\% and 30\% on average, respectively; 3) Even with higher operation latency, Lightswap-SSD still achieves 10\% -- 20\% higher throughput than Ininifswap. This is mainly because that Lightswap is co-desinged with LWT. In LWT version memcached, page fault insteads of blocking the current worker thread, it only blocks the faulting LWT. Thus other worker LWT can still get scheduled and executed by the worker thread. In contrast, in the thread version memcached, the current worker thread will be blocked once page fault happens, leading to that the CPU usage cannot be maximized.

In order to show the effectiveness of the propose paging error handling framework, we generate random UCEs and swap-in errors for address space used by memcached. Currently, we only add a simple paging error handling routine in \texttt{do\_item\_get()} of memcached. Since the item metadata and item data are placed in the same structure, and items belong to the same slab class are linked in the same LRU list. Thus, if the error handling routine finds that any item is corrupted due to paging error, it has to reset the whole slab class and returns not found for GET requests. If paging error causes any corruption in the memcached metadata, such as the hash table and slab class array, the error handling routine has to terminated memcached.


Table~\ref{tbl: paging-err} shows the results of simulated paging error handling results for Readmost workload, we generate 10 thousands swap-in errors and 15 thousands memory UCEs to memcached. As we only protect the GET operation, most of the memory UCE will cause process termination, memcached only survives in 26\% of the errors. In contrast, memcached survives in most of case (i.e., 57\%) of swap-in errors as the test workload is dominated by GET operations. We believe that with more try-catch protections, memcached can eliminate more process terminations.

\begin{table}[t]
	\centering
	\footnotesize
	\setlength{\tabcolsep}{1.1em}
	\begin{tabular}{c|cccc}
		\hline
		\bf{Error type} & \bf{Count} & \bf{Terminated (\%)} & \bf{Survived (\%)} \\
		\hline
		Swap-in error   & 10000 & 43\%    & 57\%   \\
		Memory UCE      & 15000 & 74\%    & 26\%   \\
		\hline
	\end{tabular}
	\caption{Paging error handling results. \textit{We use the OS UCE handler and the swap-int LWT to randomly generate memory UCEs and swap-in errors, respectively.}}
	\label{tbl: paging-err}
	\vspace{-0.2in}
\end{table}

\section{Related Work}
\label{sec:6}

\textbf{SSD-based swapping.} Swapping has been studied for years, with magnitudes of performance improvements compared to hard disks, SSDs based swapping becomes an attractive solution to extend the effective memory capacity. To this end, kernel-based swapping has been revisited and optimized for SSDs~\cite{FlashVM, new-linux-swap, CFLRU, os-for-hybrid, FlatFlash, speculative-paging, flashmap, mmio-flash} to enlarge the main memory. They are integrated with Linux virtual memory and rely on paging mechanism to manage the page movement between host DRAM and SSDs. Different from these application transparent approaches, runtime managed and application-aware swapping schemes~\cite{SSDAlloc, NVMalloc, ssd-hybird-memcached} are proposed to fully exploit flash's performance and alleviate the I/O amplification. However, all these swapping schemes, including both OS managed and runtime managed approaches, employ kernel-level SSD drivers and thus I/O traffics need to go through all the storage stack, which may introduce notable software overheads as the next-generation storage technology like Intel Optane~\cite{intel-optane} and KIOXIA XL-Flash~\cite{XL-Flash} are much faster than the past ones.


\textbf{Disaggregated and remote memory.} Several works~\cite{Nswap, Cashmere-VLM, global-mem-mag, remote-pager, swapping-remote-mem, page-swap-protocol, user-remote-mem} have already explored paging with remote memory instead of local SSDs, but their performance is often restricted by the slow networks and high CPU overheads. With the support of RDMA networks and emerging hardwares, it has became possible to reorganize resources into disaggregated clusters~\cite{res-disagg, net-support, disagg-mem, acc-db, mem-disagg-flx, LegoOS, tolerate-fault, Warehouse-Scale} to improve and balance the resource utilization. To achieve memory disaggregation, Fastswap~\cite{far-memory} and INFINISWAP~\cite{infiniswap} explore paging with remote memory using the kernel based swapping. FluidMem~\cite{FluidMem} supports full memory disaggregation for virtual machines through hogplug memory regions and relies userfaultfd to achieve transparent page fault handling. AIFM~\cite{AIFM} integrates swapping with application and operates at object granularity instead of page granularity to reduce network amplification. Semeru~\cite{Semeru} provides a JVM based runtime to managed applications with disaggregated memory and offloads garbage collection to servers that holding remote memory. Remote Regions~\cite{remote-regions} applies file abstraction for remote memory and provide both block (\texttt{read()/write()}) and byte (\texttt{mmap()}) access interface. In this paper, we implement a fully user-space swapping framework and co-design it with LWT for data-intensive and high-currency applications.

\textbf{Distributed share memory (DSM).} DSM systems~\cite{Munin, coherence-shared-mem, dsm-survey, Shasta, latency-tol-dsm, DSPM} provide an unified abstraction by exposing an shared global address space to applications. Different from remote memory, DSM provides an memory abstraction that data is shared across different hosts, therefore bringing significant cache coherence costs and making DSM inefficiency. To avoid the coherence costs, Partitioned Global Address Space (PGAS)~\cite{parallel-prog, perf-PGAS, PGAS-language, X10} is proposed but requires application modification. Lightswap that lets applications transparently utilize remote memory through swapping is more efficient.

\section{Conclusion}
\label{sec:7}
This paper proposes an user-space swapping mechanism that can fully exploit the high performance and low latency of emerging storage devices, as well as the RDMA-enable remote memory. We focus on three main aspects: 1) how to handle page faults in user-space effectively; 2) how to make user-space swapping both high-performance and application-transparent; 3) how to deal with paging errors which are necessary but not considered in previously works.


\bibliographystyle{unsrt}
\balance
\bibliography{lightswap}

\end{document}